\documentclass[amsmath,11pt]{article}
\usepackage{amsfonts,amsmath,amssymb,graphics,epsfig}
\usepackage{enumerate}\usepackage{theorem}

\date{\empty}

\begin{document}

\title{\bf Electromagnetic potentials in curved spacetimes}

\author{Panagiotis Mavrogiannis${}^1$ and Christos G. Tsagas${}^{1,2}$\\ {\small ${}^1$Section of Astrophysics, Astronomy and Mechanics, Department of Physics}\\ {\small Aristotle University of Thessaloniki, Thessaloniki 54124, Greece}\\ {\small ${}^2$Clare Hall, University of Cambridge, Herschel Road, Cambridge CB3 9AL, UK}}

\maketitle

\begin{abstract}
Electromagnetic potentials allow for an alternative description of the Maxwell field, the electric and magnetic components of which emerge as gradients of the vector and the scalar potential. We provide a general relativistic analysis of these potentials, by deriving their wave equations in an arbitrary Riemannian spacetime containing a generalised imperfect fluid. Some of the driving agents in the resulting wave formulae are explicitly due to the curvature of the host spacetime. Focusing on the implications of non-Euclidean geometry, we look into the linear evolution of the vector potential in Friedmann universes with nonzero spatial curvature. Our results reveal a qualitative difference in the evolution of the potential between the closed and the open Friedmann models, solely triggered by the different spatial geometry of these spacetimes. We then consider the interaction between gravitational and electromagnetic radiation and the effects of the former upon the latter. In so doing, we apply the wave formulae of both potentials to a Minkowski background and study the Weyl-Maxwell coupling at the second perturbative level. Our solutions, which apply to low-density interstellar environments away from massive compact stars, allow for the resonant amplification of both the electromagnetic potentials by gravitational-wave distortions.
\end{abstract}

\section{Introduction}\label{sI}
Although potentials are not measurable quantities themselves, they have traditionally been used in physics as an alternative to fields.\footnote{A debate regarding the physicality of the  electromagnetic potentials in quantum mechanics has been raised and driven by the interpretation efforts of the so-called Aharonov-Bohm effect.} The introduction and use of potentials is based on the principle that their differentiation leads to the realisation of the fields themselves.\footnote{There are generally more than one functional forms, the derivatives of which lead to the same result. This fact reflects the well known `gauge freedom' related to the choice of potentials.} Moreover, unless differentiation simplifies the mathematical form of a given function, potentials are generally expected to make the calculations easier to handle. In addition to  their role as auxiliary quantities that can simplify operations, it has been pointed out that electromagnetic potentials could be also introduced directly from fundamental principles (e.g.~charge conservation and action principle), as primary quantities before the fields, with the latter derived subsequently as auxiliary entities (see~\cite{LMEN,HH} for recent discussions and references therein).

One of the best known potentials in theoretical physics is that of the Maxwell field. In the context of relativity, the electromagnetic potential comes in the form of a 4-vector, with the latter comprising a timelike scalar accompanied by a 3-vector spatial component. The temporal and spatial gradients of these two entities give rise to the electric and the magnetic parts of the Maxwell field. In principle, one could use either the potential or the field description when studying electromagnetic phenomena. Most of the available work, however, employs the latter rather than the former. As a result, certain aspects regarding the behaviour of the electromagnetic potentials are still missing from the available literature. One of the relatively less explored aspects is the coupling between the Maxwell potentials and the geometry of their host spacetime. This area also appears to be one of the most challenging, since the study of electromagnetism in curved spaces has led to rather unconventional phenomenology in a number of occasions (e.g.~see~\cite{DWB}-\cite{PM} for a representative though incomplete list)

One would like to know, in particular, whether and how the Ricci and the Weyl curvature, namely the local and the far components of the gravitational field, couple to the electromagnetic potentials. The interaction with the intrinsic and the extrinsic curvature of the 3-dimensional space hosting the Maxwell field and its implications are additional questions as well. This study attempts to shed light upon these matters, by providing the first (to the best of our knowledge) general relativistic wave-equations of both the vector and the scalar electromagnetic potentials in an arbitrary Riemannian spacetime containing a general imperfect fluid. In so doing, we pay special attention to the geometrical features of the emerging wave formulae and more specifically to that of the vector potential. These features result from the geometrical interpretation of gravity that Einstein's theory introduces and from the vector nature of the Maxwell field. The most compact expression, reflecting the aforementioned coupling between electromagnetism and spacetime curvature, is perhaps the wave equation of the 4-potential. In the absence of sources, the latter reads
\begin{equation}
\left(\square\delta^a{}_b- R^a{}_b\right)A_a= 0\,,  \label{D'Alen}
\end{equation}
where $\square$ is the d' Alembertian operator, $\delta^a{}_b$ the Kronecker delta, $R^a{}_b$ is the Ricci curvature tensor and $A_a$ is the electromagnetic 4-potential~\cite{MTW}-\cite{DFC}. Note that $\square\delta^a{}_b- R^a{}_b$ defines the so-called de Rham wave-operator, which acts as the generalised d' Alembertian in curved spacetimes.\footnote{Following~\cite{W}, the Ricci curvature term on the left-hand side of (\ref{D'Alen}) underlines the particular attention one needs to pay when obtaining the general relativistic electromagnetic equations from their (flat space) special relativistic counterparts by means of the so-called ``minimal substitution rule''.}

The evolution of the electric and the magnetic components of the Maxwell field in a general spacetime was studied in~\cite{T1}, by means of the 1+3 covariant approach to general relativity~\cite{TCM,EMM}, with the propagation equations given in the form of wave-like formulae. Typically, these describe forced oscillations traveling at the speed of light, with driving terms that include, among others, the curvature of the host spacetime~\cite{T1}. Here, we provide an alternative (though still 1+3 covariant) treatment involving the electromagnetic potential, instead of the fields themselves. Starting from Maxwell's equations and adopting the Lorenz gauge, we arrive at a pair of wave-like formulae for the vector and the scalar potentials of the Maxwell field. These hold in any Riemannian spacetime, just like the ones for the electric and the magnetic fields obtained in~\cite{T1}. The qualitative difference is that, there, the matter component was represented by a perfect fluid, whereas here it has the form of a generalised imperfect medium.

The wave equations derived in~\cite{T1} were linearised around a Friedmann-Robertson-Walker (FRW) cosmology with nonzero spatial curvature. In the absence of charges, the solutions revealed that spatial-curvature effects could enhance the amplitude of electromagnetic waves in Friedmann models with hyperbolic spatial geometry. Here, we provide an alternative treatment involving the electromagnetic potential, instead of the actual fields. By linearising the wave-like formula of the vector potential around a Friedmann universe with non-Euclidean spatial hypersurfaces, we arrive at analytic solutions representing generalised forced oscillations. The frequency and the amplitude of the latter depend on the spatial geometry of the background FRW model. The amplitude of the vector potential, in particular, is enhanced when the background Friedmann universe has negative 3-curvature (in full agreement with~\cite{T1} -- see also~\cite{BT}).

We then consider the coupling between the Weyl and the Maxwell fields, namely the interaction between gravitational and electromagnetic radiation. More specifically, we analyse the effects of the former on the propagation of the latter. Adopting the Minkowski space as our background, we consider the aforementioned Weyl-Maxwell coupling at the second perturbative order. Employing the wave-like formulae of the vector and the scalar potentials, we find that the interaction between gravitational and electromagnetic waves can lead to the resonant amplification of the latter. In most realistic physical situations, these resonances should result into the linear growth of the electromagnetic signal,  in agreement with the earlier studies of~\cite{T3,KT}. Note that, given the Minkowski nature of the adopted background spacetime, our analysis and our results apply away from massive compact stars to the low-density interstellar/intergalactic environments, where the gravitational field is expected to be weak.

The manuscript starts with a brief presentation of electromagnetic fields and potentials in the framework of the 1+3 covariant formalism, with emphasis on the coupling between electromagnetism and spacetime geometry. The wave-formulae for the vector and the scalar potentials are extracted from Maxwell's equations in~\S~\ref{sMEWF}. These are subsequently linearised around a FRW background with nonzero spatial curvature in~\S~\ref{ssEMPFRWSs} and then employed to study the Weyl-Maxwell coupling at second order in~\S~\ref{ssG-WEEMSs}. We also note that the interested reader can find the necessary background information on the 1+3 covariant formalism in Appendix~\ref{AppA}. Additional technical details and guidance for reproducing the wave formulae of the vector and the scalar electromagnetic potentials are given in Appendix~\ref{AppB}.

\section{Electromagnetic fields and potentials}\label{sEMFPs}
Maxwell's equations allow for the existence of the electromagnetic 4-potential, the gradients of which lead to the electric and magnetic fields measured by the observers. Therefore, depending on the problem in hand, one is free to choose either representation of the Maxwell field.

\subsection{Electric and magnetic vectors}\label{ssEMVs}
The electromagnetic field is covariantly described by the antisymmetric Faraday tensor ($F_{ab}=F_{[ab]}$). Relative to a family of observers, moving along timelike worldlines tangent to the 4-velocity $u_a$ (with $u_au^a=-1$ -- see also Appendix~\ref{AppA1}), the Faraday tensor splits as
\begin{equation}
F_{ab}= 2u_{[a}E_{b]}+ \epsilon_{abc}B^{c}\,,  \label{Fab1}
\end{equation}
with $E_a=F_{ab}u^b$ and $B_a=\epsilon_{abc}F^{bc}/2$ representing its electric and magnetic components. Note that $\epsilon_{abc}= \epsilon_{[abc]}$ is the Levi Civita tensor of the 3-D space orthogonal to the observers' worldlines (i.e.~$\epsilon_{abc}u^c=0$ -- see Appendix~\ref{AppA2} as well). The latter guarantees that both the electric and the magnetic fields are 3-D spacelike vectors with $E_au^a=0=B_au^a$.

The Faraday tensor also determines the electromagnetic energy-momentum tensor, which satisfies the covariant expression
\begin{equation}
T^{(em)}_{ab}= -F_{ac}F^c{}_b- \frac{1}{4}\,F_{cd}F^{cd}g_{ab}\,,  \label{Tem}
\end{equation}
where $g_{ab}$ is the spacetime metric with signature $(-,+,+,+)$. Equivalently, we may use decomposition  (\ref{Fab1}) to write
\begin{equation}
T^{(em)}_{ab}= \frac{1}{2}\left(E^{2}+B^{2}\right)u_{a}u_{b}+ \frac{1}{6}\left(E^{2}+B^{2}\right)h_{ab}+ 2\mathcal{Q}_{(a}u_{b)}+ \Pi_{ab}\,, \label{eqn:T-EM1}
\end{equation}
thus explicitly involving the electric and magnetic components. In the above, $E^2=E_aE^a$, $B^2=B_aB^a$ are the squared magnitudes of the electric and the magnetic fields respectively, while $h_{ab}=g_{ab}+u_au_b$ is the symmetric tensor that projects orthogonal to the $u_a$-field (with $h_{ab}u^b=0$ -- see Appendix~\ref{AppA1}).\footnote{It is worth comparing Eq.~(\ref{eqn:T-EM1}) to the stress-energy tensor of a general imperfect fluid. Decomposed relative to the $u_a$ 4-velocity field, the latter reads
\begin{equation}
T^{(m)}_{ab}= \rho u_au_b+ ph_{ab}+ 2q_{(a}u_{b)}+ \pi_{ab}\,,  \label{mTab}
\end{equation}
with $\rho=T_{ab}u^au^b$ and $p=T_{ab}h^{ab}/3$ representing the energy density and the (isotropic) pressure of the matter, while $q_a=-h_a{}^bT_{bc}u^c$ and $\pi_{ab}=h_{\langle a}{}^ch_{b\rangle}{}^dT_{cd}$ are the associated energy-flux vector and viscosity tensor respectively (with $q_au^a=0$, $\pi_{ab}=\pi_{(ab)}$ and $\pi_{ab}u^b=0=\pi_a{}^a$).} Expression (\ref{eqn:T-EM1}) allows for an imperfect-fluid description of the electromagnetic field, where $\rho^{(em)}=(E^2+B^2)/2$ is the energy density and $p^{(em)}=(E^2+B^2)/6$ is the isotropic pressure. These are supplemented by an effective energy-flux, given by the Poynting vector $\mathcal{Q}_a=\epsilon_{abc}E^bB^c$, and by the electromagnetic viscosity tensor $\Pi_{ab}=-E_{\langle a}E_{b\rangle}-B_{\langle a}B_{b\rangle}$ (with $\mathcal{Q}_au^a=0=\Pi_{ab}u^b$ by construction).\footnote{Round and square brackets indicate symmetrisation and antisymmetrisation as usual. Angular brackets, on the other hand, denote the symmetric and traceless part of spacelike tensors. In particular, $E_{\langle a}E_{b\rangle}=E_{(a}E_{b)}-(E^2/3)h_{ab}$. Note that angled brackets are also used to denote the orthogonally projected part of vectors. For instance, $\dot{\mathcal{A}}_{\langle a\rangle}= h_a{}^b\dot{\mathcal{A}}_b$ (see Eq.~(\ref{el-field-pot}) in \S~\ref{ssVSPs} next).}

\subsection{Vector and scalar potentials}\label{ssVSPs}
The Faraday tensor can be also expressed in terms of the electromagnetic 4-potential, the existence of which is implied by the second of Maxwell's equations (see (\ref{eqn:Maxw-eqns1}b) in \S~\ref{ssMEs} below). More specifically, we have
\begin{equation}
F_{ab}= \nabla_aA_b- \nabla_bA_a= \partial_aA_b- \partial_bA_a\,,  \label{Fab2}
\end{equation}
where the second equality follows from the symmetry of the connection in Riemannian spaces. Relative to the $u_a$-field of the timelike observers, the 4-potential splits into its temporal and spatial parts according to the decomposition
\begin{equation}
A_a= \Phi u_a+ \mathcal{A}_a\,,  \label{Aa}
\end{equation}
with $\Phi=-A_au^a$ being the scalar potential and $\mathcal{A}_a=h_a{}^bA_b$ representing the 3-vector potential (so that $\mathcal{A}_au^a=0$). Setting the divergence of the above to zero leads to the expression
\begin{equation}
\dot{\Phi}+ \Theta\Phi+{\rm D}^{a}\mathcal{A}_{a}+\dot{u}^{a}\mathcal{A}_{a}=0\,, \label{Lorenz-gauge}
\end{equation}
which reproduces the familiar Lorenz-gauge condition (i.e.~$\nabla^{a}A_{a}=0$) in 1+3 covariant form. Note that $\dot{\Phi}=u^a\nabla_a\Phi$ is the time-derivative of the scalar potential, relative to the $u_a$-field. Also, the variables $\dot{u}_a=u^b\nabla_bu_a$ (with $\dot{u}_au^a=0$ by construction) and $\Theta={\rm D}^au_a=\nabla^au_a$ are (irreducible) kinematic quantities, respectively representing the 4-acceleration vector and the volume scalar associated with the 4-velocity field (see Appendices~\ref{AppA1} and~\ref{AppA2} for more details).

Combining the definitions of the electric and the magnetic fields (see Eq.~(\ref{Fab1}) in \S~\ref{ssEMVs} above) with decomposition (\ref{Aa}), leads to
\begin{equation}
E_a= -\dot{\mathcal{A}}_{\langle a\rangle}- \frac{1}{3}\,\Theta\mathcal{A}_a- \left(\sigma_{ab}-\omega_{ab}\right)\mathcal{A}^b- {\rm D}_{a}\Phi- \Phi\dot{u}_a  \label{el-field-pot}
\end{equation}
and
\begin{equation}
B_a= \text{curl}\mathcal{A}_a- 2\Phi\omega_a\,,
\label{magn-field-pot}
\end{equation}
recalling that $\dot{\mathcal{A}}_{\langle a\rangle}= h_a{}^b\dot{\mathcal{A}}_b$ (see footnote~5). Also, $\sigma_{ab}={\rm D}_{\langle b}u_{a\rangle}$ and $\omega_{ab}={\rm D}_{[b}u_{a]}$ are the shear and the vorticity tensors respectively (with $\sigma_{ab}u^b=0=\omega_{ab}u^b$ by default). These quantities, together with the 4-acceleration vector and the volume scalar defined previously, completely determine the kinematics of the observers' worldlines (see also Appendix~\ref{AppA2}). In addition, we have ${\rm curl}\mathcal{A}_a= \epsilon_{abc}{\rm D}^b\mathcal{A}^c$ and $\omega_a=\epsilon_{abc}\omega^{bc}/2$ (with $\omega_au^a=0$) by construction. Relations (\ref{el-field-pot}) and (\ref{magn-field-pot}) express both components of the Maxwell field in terms of the scalar and the 3-vector potentials. Note the kinematic terms on the right-hand side of these expressions, which are induced by the relative motion of the neighbouring observers.\footnote{Very similar, though not entirely 1+3 covariant, expressions for the electric and magnetic fields in terms of the potentials can be found in~\cite{TMcD}. Also, written in the static Minkowski space, where $\dot{u}_{a}=0=\Theta= \sigma_{ab}=\omega_{ab}$ by default, Eqs.~\eqref{el-field-pot} and~\eqref{magn-field-pot} recast into the more familiar expressions (e.g.~see~\cite{J})
\begin{equation}
E_{a}= -\partial_{t}\mathcal{A}_{a}- \partial_{a}\Phi \hspace{15mm} {\rm and} \hspace{15mm}
B_{a}= \text{curl}\mathcal{A}_{a}\,,  \label{Minkowski}
\end{equation}
respectively (with $\text{curl}\mathcal{A}_{a}= \epsilon_{abc}\partial^{b}\mathcal{A}^{c}$ in this case).}

\section{Electromagnetism and spacetime geometry}\label{EMSG}
The electromagnetic field is the only known energy source of vector nature. This feature facilitates a dual coupling between the Maxwell field and the geometry of the host spacetime, mediated by Einstein's equations, on the one hand, and by the Ricci identities on the other.

\subsection{Electromagnetic fields and curvature}\label{ssEMFC}
Electromagnetic fields are carriers of energy and, as such, they couple to the curvature of their host spacetime through the Einstein equations
\begin{equation}
R_{ab}- \frac{1}{2}\,Rg_{ab}= \kappa\left(T^{(m)}_{ab}+ T^{(em)}_{ab}\right)\,,  \label{EFE}
\end{equation}
where $R_{ab}$ and $R=R_a{}^a$ are the Ricci tensor and the Ricci scalar respectively, while $g_{ab}$ is the spacetime metric. Also, $T^{(m)}_{ab}$ is the energy-momentum tensor of the matter, $T^{(em)}_{ab}$ is that of the electromagnetic field (see Eqs.~(\ref{Tem}) and \eqref{eqn:T-EM1} in \S~\ref{ssEMVs} earlier) and $\kappa=8\pi G$ (with $c=1$) is the gravitational constant. Given their vector nature, the electric and the magnetic fields have an additional (purely geometrical) interaction with the spacetime curvature via the Ricci identities
\begin{equation}
2\nabla_{[a}\nabla_{b]}E_{c}= R_{abcd}E^{d}\,, \label{eqn:Ricc-Electr}
\end{equation}
written here for the electric field ($E_{a}$), with $R_{abcd}$ representing the Riemann (curvature) tensor of the 4-dimensional spacetime. Projecting \eqref{eqn:Ricc-Electr} into the observers' 3-D rest-space leads to the 3-Ricci identities. The latter read~\cite{TCM,EMM}
\begin{equation}
2{\rm D}_{[a}{\rm D}_{b]}E_{c}= -2\omega_{ab}\dot{E}_{\langle c\rangle}+ \mathcal{R}_{dcba}E^{d}\,,  \label{3-D-Ricci-EL}
\end{equation}
with $\mathcal{R}_{abcd}$ representing the 3-D Riemann tensor that determines the geometry of the spatial sections (see Appendix~\ref{AppA3} for details). It goes without saying that relations exactly analogous to (\ref{eqn:Ricc-Electr}) and (\ref{3-D-Ricci-EL}) hold for the magnetic vector as well.

\subsection{Electromagnetic potentials and curvature}\label{ssEMPC}
The coupling of the Maxwell field to the Riemann tensor of the whole spacetime, as well as to the curvature of the host 3-space (see Eqs.~(\ref{eqn:Ricc-Electr}) and (\ref{3-D-Ricci-EL}) respectively), is a purely geometrical interaction that goes beyond the usual interplay between matter and geometry monitored by Einstein's equations (see (\ref{EFE}) above). Not surprisingly, an analogous interaction also holds between the electromagnetic potential and the curvature of the host space. More specifically, applying the 4-Ricci identities to the 4-potential, gives
\begin{equation}
2\nabla_{[a}\nabla_{b]}A_{c}= R_{abcd}A^{d}\,. \label{eqn:Ricci-id-4pot}
\end{equation}
The scalar potential and the vector potential, on the other hand, satisfy the following versions of the 3-Ricci identities, that is
\begin{equation}
{\rm D}_{[a}{\rm D}_{b]}\Phi= -\dot{\Phi}\omega_{ab} \label{Ricci-scalar-pot}
\end{equation}
and
\begin{equation}
2{\rm D}_{[a}{\rm D}_{b]}\mathcal{A}_{c}= -2\omega_{ab}\dot{\mathcal{A}}_{\langle c\rangle}+ \mathcal{R}_{dcba}\mathcal{A}^{d}\,,  \label{Ricci-vec-pot}
\end{equation}
respectively. Note that, according to Eq.~(\ref{Ricci-scalar-pot}), the spatial gradients of scalars do not commute in rotating spacetimes. This result, as well as the vorticity terms on the right-hand side of (\ref{3-D-Ricci-EL}) and (\ref{Ricci-vec-pot}), are direct consequences of the Frobenius theorem, which ensures that rotating spacetimes do not contain integrable spacelike hypersurfaces (e.g.~see~\cite{EBH} for a discussion).

\section{Maxwell's equations and wave formulae}\label{sMEWF}
Starting form Maxwell's equations one can extract a set of generalised wave formulae. Written in their full form, the latter monitor the evolution of the electric and the magnetic fields in an arbitrary Riemannian spacetime~\cite{T1}. In the current section, we will derive the corresponding wave-like equations of the electromagnetic vector and scalar potentials.

\subsection{Maxwell's equations}\label{ssMEs}
The behaviour of the electromagnetic fields is determined by Maxwell equations. Written in terms of the Faraday tensor, the latter take the compact covariant form
\begin{equation}
\nabla^{b}F_{ab}= J_{a}\hspace{15mm} \text{and} \hspace{15mm} \nabla_{[c}F_{ab]}= 0\,,
\label{eqn:Maxw-eqns1}
\end{equation}
where $J_a=\mu u_a+\mathcal{J}_a$ is the electric 4-current (with $\mu=-J_au^a$ and $\mathcal{J}_a= h_a{}^bJ_b$ representing the electric charge and the associated 3-current respectively). Note that constraint (\ref{eqn:Maxw-eqns1}b) allows for the existence of the electromagnetic 4-potential ($A_{a}$) seen in Eq.~(\ref{Fab2}).

Projecting Maxwell's formulae orthogonally to the $u_{a}$-field provides the propagation equations of the electric and the magnetic fields, namely (e.g.~see~\cite{TCM,EMM})
\begin{equation}
\dot{E}_{\langle a\rangle}= -\frac{2}{3}\Theta E_a+ \left(\sigma_{ab}+\epsilon_{abc}\omega^c\right)E^b+ \epsilon_{abc}\dot{u}^bB^c+ {\rm curl}B_a- \mathcal{J}_a \label{eqn:el-field-prop}
\end{equation}
and
\begin{equation}
\dot{B}_{\langle a\rangle}= -\frac{2}{3}\Theta B_a+ \left(\sigma_{ab}+\epsilon_{abc}\omega^c\right)B^b- \epsilon_{abc}\dot{u}^bE^c- {\rm curl}E_a\,, \label{eqn:magn-field-prop}
\end{equation}
respectively. Projecting (\ref{eqn:Maxw-eqns1}a)  and (\ref{eqn:Maxw-eqns1}b) along $u_{a}$, on the other hand, leads to the constraints
\begin{equation}
{\rm D}^aE_a= \mu- 2\omega^aB_a \hspace{15mm} {\rm and} \hspace{15mm} {\rm D}^aB_a= 2\omega^aE_a\,.  \label{magn-div}
\end{equation}
All the kinematic terms seen on the right-hand side of (\ref{eqn:el-field-prop})-(\ref{magn-div}b) are induced by the relative motion of the neighbouring observers and vanish in a static Minkowski-like spacetime. Recall that analogous relative-motion effects were also observed in Eqs.~(\ref{el-field-pot}) and (\ref{magn-field-pot}) in~\S~\ref{ssVSPs} earlier. Finally, we should note that expressions \eqref{eqn:el-field-prop}, \eqref{eqn:magn-field-prop}, (\ref{magn-div}a) and (\ref{magn-div}b) constitute 1+3 covariant versions of Amp\`ere's, Faraday's, Coulomb's and Gauss's laws respectively.

\subsection{Wave equations for the potentials}\label{ssWEPs}
The wave formulae of both electromagnetic potentials follow from the first of Maxwell's equations (see expression (\ref{eqn:Maxw-eqns1}a) in \S~\ref{ssMEs}). More specifically, the wave formula of the vector potential follows from Amp\`ere's law, whereas that of its scalar counterpart derives from Coulomb's law (see Eqs.~(\ref{eqn:el-field-prop}) and (\ref{magn-div}a) respectively). Next, we will provide the corresponding expressions and refer the reader to Appendix~\ref{AppB} for the technical details. Note that both formulae hold in an arbitrary Riemannian spacetime filled with a general imperfect fluid.

Substituting expressions \eqref{el-field-pot} and \eqref{magn-field-pot} into Eq.~(\ref{eqn:el-field-prop}) and then imposing the Lorenz-gauge condition (see constraint (\ref{Lorenz-gauge}) in \S~\ref{ssVSPs}, as well as Appendix~\ref{AppB1} for additional details, auxiliary relations and intermediate steps), we arrive at
\begin{eqnarray}
\ddot{\mathcal{A}}_{\langle a\rangle}- {\rm D}^{2}\mathcal{A}_{a}&=& -\Theta\dot{\mathcal{A}}_{\langle a\rangle}+ {1\over3}\left[\frac{1}{2}\,\kappa\left(\rho+3p\right) -\frac{1}{3}\,\Theta^2+ 4\sigma^2- {1\over3}\,\dot{u}_b\dot{u}^b\right]\mathcal{A}_a- \mathcal{R}_{ba}\mathcal{A}^b+ E_{ab}\mathcal{A}^b \nonumber\\ &&\left[{1\over3}\,\Theta(\sigma_{ab}+\omega_{ab}) -{1\over2}\,\kappa\pi_{ab}+2\sigma_{c\langle a}(\sigma^c{}_{b\rangle}-\omega^c{}_{b\rangle}) -2\sigma_{c[a}\omega^c{}_{b]}-{1\over3}\,\dot{u}_{\langle a}\dot{u}_{b\rangle}\right]\mathcal{A}^b \nonumber\\ && -{7\over3}\,\dot{\Phi}\dot{u}_a-\left[\ddot{u}_{\langle a\rangle} +\Theta\dot{u}_a-(\sigma_{ab}+3\omega_{ab})\dot{u}^b-{\rm D}_a\Theta -2{\rm curl}\omega_a\right]\Phi \nonumber\\ &&+\dot{u}^b\left({\rm D}_{\langle b}\mathcal{A}_{a\rangle}+{\rm D}_{[b}\mathcal{A}_{a]}\right)+ {2\over3}\,\Theta{\rm D}_a\Phi+ 2(\sigma_{ab}+\omega_{ab}){\rm D}^b\Phi+ \mathcal{J}_a\,,  \label{eqn:vec-pot}
\end{eqnarray}
where ${\rm D}^2={\rm D}^a{\rm D}_a$ is the spatial covariant Laplacian operator. The above is a wave-like formula with extra terms, which reflect the fact that the host spacetime is not static, it contains matter and has non-Euclidean geometry. The latter, namely gravity, is represented by the Ricci tensor of the spatial sections and by the electric Weyl tensor ($\mathcal{R}_{ab}$ and $E_{ab}$ respectively -- see Appendix~\ref{AppA3} for details). The explicit presence of the 3-Ricci tensor and of the electric Weyl tensor on the right-hand side of (\ref{eqn:vec-pot}) ensures that spatial curvature and the Weyl field can drive fluctuations in the vector potential. Both terms are the direct result of the gravito-electromagnetic coupling reflected in Ricci identities (see Eqs.~(\ref{eqn:Ricci-id-4pot}) and (\ref{Ricci-vec-pot}) in \S~\ref{ssEMPC}).

Proceeding in an analogous way, one also obtains the wave equation of the scalar potential. The latter follows from Coulomb's law (see expression (\ref{magn-div}a) in \S~\ref{ssMEs} above) after making use of Eqs.~(\ref{el-field-pot}) and (\ref{magn-field-pot}). In so doing (see Appendix~\ref{AppB2} for further details), one arrives at
\begin{eqnarray}
\ddot{\Phi}-{\rm D}^{2}\Phi&=& -\frac{5}{3}\,\Theta\dot{\Phi}+ \dot{u}^{a}{\rm D}_{a}\Phi+ \left[\frac{1}{2}\,\kappa\left(\rho+3p\right) -\frac{1}{3}\,\Theta^2+2\left(\sigma^2+\omega^2\right) -\dot{u}^{a}\dot{u}_{a}\right]\Phi \nonumber\\
&&+\left[{\rm D}_{a}\Theta-\frac{4}{3}\,\Theta\dot{u}_a +2\text{curl}\omega_a-2\kappa q_{a} +\sigma_{ab}\dot{u}^b+3\epsilon_{abc}\dot{u}^b\omega^c -\ddot{u}_a\right]\mathcal{A}^a \nonumber\\
&&+2\sigma_{ab}{\rm D}^b\mathcal{A}^a- 2\omega_a{\rm curl}\mathcal{A}^a- 2\dot{u}_a\dot{\mathcal{A}}^a+ \mu\,,
\label{eqn:scalar-pot}
\end{eqnarray}
which (like Eq.~(\ref{eqn:vec-pot}) above) shows wave-propagation at the speed of light. Comparing the above to expression (\ref{eqn:vec-pot}), one immediately notices the complete absence of explicit curvature terms on the right-hand side of (\ref{eqn:scalar-pot}). This difference, which was largely anticipated given the scalar nature of the related potential, implies that spacetime curvature can only indirectly affect the evolution of $\Phi$.\footnote{An alternative way of extracting the wave-like equations \eqref{eqn:vec-pot} and \eqref{eqn:scalar-pot} is by substituting (\ref{Fab2}) into Maxwell's formulae (see Eqs.~(\ref{eqn:Maxw-eqns1}) in \S~\ref{ssMEs}) and then projecting the resulting expression along and orthogonal to the $u_{a}$-field.}

Unlike the first of Maxwell's formulae, the second (see Eq.~(\ref{eqn:Maxw-eqns1}b) in \S~\ref{ssMEs}) is trivially satisfied by the electromagnetic potential. Recall that expression (\ref{eqn:Maxw-eqns1}b) is the one that allows for the presence of the potential in the first place. As a result, substituting relations (\ref{el-field-pot}) and (\ref{magn-field-pot}) into Faraday's and Gauss' laws (see Eqs.~(\ref{eqn:magn-field-prop}) and (\ref{magn-div}b) in \S~\ref{ssMEs}) does not provide any additional propagation or constraint equations, but instead leads to trivial identities.

Finally, before closing this section, we should note that the matter terms seen on the right-hand side of (\ref{eqn:vec-pot}) and (\ref{eqn:scalar-pot}) correspond to the total (effective) fluid. Put another way, the energy density ($\rho$), the pressure ($p$), the energy flux ($q_a$) and the viscosity ($\pi_{ab}$) contain the corresponding contributions of the electromagnetic field (see \S~\ref{ssEMVs} previously) as well.

\section{Application to cosmology and astrophysics}\label{sACAs}
The wave equations of the previous section hold in a general Riemannian spacetime containing an imperfect fluid in the presence of the electromagnetic field. As a result, they  can be linearised around almost any background model and applied to a variety of astrophysical and cosmological environments. In what follows, we will consider two characteristic applications.

\subsection{Electromagnetic potentials in FRW 
spacetimes}\label{ssEMPFRWSs}
The high symmetry of the Friedmann models, namely their spatial homogeneity and isotropy ensure that they cannot narurally accommodate electromagnetic fields. Therefore, in order to study the Maxwell field in FRW-like environments, one needs to introduce it as a perturbation.\footnote{One may also introduce a sufficiently random electromagnetic field, which does not affect the isotropy of the Friemann host but only adds to the total energy of the matter sources. In other words, one may assume that $\langle E_a\rangle=0=\langle B_a\rangle$ on average, whereas $\langle E^2\rangle$, $\langle B^2\rangle\neq0$ (e.g.~see~\cite{LKNS} and references therein). This approach does not serve the purposes of our study, however, given that the effects we are primarily interested in stem from the vector nature of the Maxwell field.} Proceeding along these lines, let us consider an FRW universe filled with a single perfect fluid and then perturb it by allowing for the presence of a source-free electromagnetic field. Then, the wave formula of the vector potential (see Eq.~(\ref{eqn:vec-pot}) in \S~\ref{ssWEPs} above) linearises to
\begin{equation}
\ddot{\mathcal{A}}_a- {\rm D}^2\mathcal{A}_a= -\Theta\dot{\mathcal{A}}_a- \frac{1}{9}\,\Theta^2\mathcal{A}_a+ \frac{1}{6}\,\kappa\left(\rho+3p\right)\mathcal{A}_a- \mathcal{R}_{ab}\mathcal{A}^b\,,  \label{eqn:vec-pot-FRW}
\end{equation}
given that $\mathcal{J}_a=0$ in the absence of charges. Note that $\mathcal{R}_{ab}=(2K/a^{2})h_{ab}$ in a Friedmann universe, with $K=0,\pm1$ being the 3-curvature index. Starting from the above, recalling that $\Theta=3H$ in FRW cosmologies and employing the background Raychaudhuri equation, namely
\begin{equation}
\dot{H}= -H^2- \frac{1}{6}\,\kappa\left(\rho+3p\right)\,,  \label{Ray}
\end{equation}
the linear relation \eqref{eqn:vec-pot-FRW} recasts into
\begin{equation}
\ddot{\mathcal{A}}_a- {\rm D}^2\mathcal{A}_a= -3H\dot{\mathcal{A}}_a- \left(2H^2+\dot{H}+\frac{2K}{a^2}\right)\mathcal{A}_{a}\,, \label{eqn:vec-pot-FRW2}
\end{equation}
where $a=a(t)$ is the cosmological scale factor (with $H=\dot{a}/a$ -- see also Appendix~\ref{AppA2}). The above is a wave-like differential equation, with extra terms due to the expansion and gravity and with time-dependent coefficients. After a simple Fourier decomposition, Eq.~(\ref{eqn:vec-pot-FRW2}) leads to the following expression
\begin{equation}
\ddot{\mathcal{A}}_{(n)}+ \left(\frac{n}{a}\right)^2\mathcal{A}_{(n)}= -3H\dot{\mathcal{A}}_{(n)}- \left(2H^2+\dot{H}+\frac{2K}{a^2}\right)\mathcal{A}_{(n)}\,,  \label{hcAa1}
\end{equation}
for the $n$-th harmonic mode of the vector potential.\footnote{We employ the familiar Fourier expansion $\mathcal{A}_a= \sum_{n}\mathcal{A}_{(n)}\mathcal{Q}^{(n)}_a$, in terms of the vector harmonics $\mathcal{Q}^{(n)}_a$, so that ${\rm D}_{a}\mathcal{A}_{(n)}=0= \dot{\mathcal{Q}}^{(n)}_a$ and ${\rm D}^{2}\mathcal{Q}^{(n)}_a= -(n/a)^2\mathcal{Q}^{(n)}_a$. Note that $n$ is the Laplacian eigenvalue, which coincides with the comoving wavenumber of the mode when the background FRW universe is spatially flat. In that case, as well as in Friedmann models with hyperbolic spatial sections (i.e.~when $K=0,-1$) the eigenvalue is continuous with $n^2>0$. When $K=+1$, on the other hand, the eigenvalue is discrete with $n^2\geq3$ (e.g.~see~\cite{TCM,EMM}).}

Our next step is to recast (\ref{hcAa1}) with respect to the conformal time ($\eta= \int(dt/a)=\int da/(\dot{a}a)$). In so doing, we arrive at
\begin{equation}
\mathcal{A''}_{(n)}+ n^{2}\mathcal{A}_{(n)}= -2\left(\frac{a'}{a}\right)\mathcal{A'}_{(n)}- \left(\frac{a''}{a}+2K\right)\mathcal{A}_{(n)}\,,  \label{hcAa2}
\end{equation}
with the primes indicating differentiation in terms of the conformal time. Finally, after introducing the rescaled potential $\mathfrak{A}_{(n)}=a\mathcal{A}_{(n)}$, the above takes the compact form
\begin{equation}
\mathfrak{A}''_{(n)}+ \left(2K+n^{2}\right)\mathfrak{A}_{(n)}= 0\,,  \label{hcAa3}
\end{equation}
where $K=0,\pm1$ depending on the geometry of the background spatial hypersurfaces. This wave equation agrees with the one obtained in~\cite{K} (compare to Eq.~(17) there), provided the latter is applied to source-free electromagnetic fields, or to environments of very low (essentially zero) electrical conductivity.

Assuming FRW backgrounds with Euclidean or spherical spatial geometry, namely setting $K=0,+1$, Eq.~(\ref{hcAa3}) leads to the following oscillatory solution for the vector potential
\begin{equation}
\mathcal{A}_{(n)}= \frac{1}{a}\left[\mathcal{C}_1\cos\left(\sqrt{n^2+2K}\,\eta\right) +\mathcal{C}_2\sin\left(\sqrt{n^2+2K}\,\eta\right)\right]\,,  \label{Aa0+1}
\end{equation}
with the integration constants ($\mathcal{C}_1$ and $\mathcal{C}_2$) determined by the initial conditions. Therefore, in flat and closed Friedmann models, the vector potential oscillates with amplitude that decays as $\mathcal{A}_{(n)}\propto1/a$ on all scales. Recall that, in a flat FRW universe, the wavenumber is continuous with $n^2>0$, while $\eta>0$ as well. When dealing with closed FRW models, on the other hand, $n$ is discrete with $n^2\geq3$. Also, in that case, the conformal time satisfies the constraint $\eta\in[0,2\pi/(1+3w)]$, where $w=p/\rho$ is the barotropic index of the matter.

In spatially open Friedmann universes, with $K=-1$, the coefficient $2K+n^2$ of the second term on the left-hand side of (\ref{hcAa3}) is positive only when $n^2>2$. On the corresponding scales, the vector potential oscillates with decreasing amplitude in line with solution (\ref{Aa0+1}). However, on longer wavelengths (those with $0<n^2<2$) the solution of Eq.~(\ref{hcAa3}) reads
\begin{eqnarray}
\mathcal{A}_{(n)}&=& \frac{1}{a}\left[\mathcal{C}_1\cosh\left(\sqrt{|n^2+2K|}\,\eta\right) +\mathcal{C}_2\sinh\left(\sqrt{|n^2+2K|}\,\eta\right)\right] \nonumber\\ &=&\frac{1}{a}\left(\mathcal{C}_3e^{\eta\sqrt{2-n^2}} +\mathcal{C}_4e^{-\eta\sqrt{2-n^2}}\right)\,,  \label{Aa-1}
\end{eqnarray}
since $K=-1$. In Friedmann models with hyperbolic spatial geometry the scale factor evolves as $a\propto\sinh(\beta\eta)^{1/\beta}$, where $\beta=(1+3w)/2$ and $\eta>0$. Therefore, when the universe is dominated by conventional matter with $\beta>0$, the late-time evolution of the scale factor is $a\propto e^{\eta}$. In such an environment, the dominant mode of solution (\ref{Aa-1}) evolves according to the power law
\begin{equation}
\mathcal{A}_{(n)}\propto a^{\sqrt{2-n^2}-1}\,,  \label{amplAa}
\end{equation}
with $0<n^2<2$. Consequently, as long as $1<n^2<2$ the vector potential keeps decaying, though at a rate slower than $\mathcal{A}\propto1/a$. However, on longer wavelengths (those with $0<n^2<1$), the amplitude of the vector potential starts increasing. In fact, at the infinite wavelength limit (i.e.~for $n\rightarrow0$) the vector potential grows as $\mathcal{A}_{(0)}\propto a^{\sqrt{2}-1}$. Therefore, in perturbed FRW cosmologies with open spatial sections and on sufficiently large scales, the decay of the electromagnetic vector potential is reversed solely due to curvature effects.

Solutions (\ref{Aa0+1}) and (\ref{Aa-1}) are in full agreement with the ones describing the linear evolution of electric and magnetic fields in Friedmann models (see~\cite{T1} for details). When the FRW background is flat or closed (i.e.~for $K=0,+1$), for example, the magnetic field obeys the oscillatory solution
\begin{equation}
B_{(n)}= \frac{1}{a^2}\left[\mathcal{C}_1\cos\left(\sqrt{n^2+2K}\,\eta\right) +\mathcal{C}_2\sin\left(\sqrt{n^2+2K}\,\eta\right)\right]\,, \label{eqn:B-amplitude1}
\end{equation}
on all scales. The above result also holds in perturbed Friedmann cosmologies with open spatial sections, as long as $n^2>2$. Otherwise, namely on longer wavelengths with $0<n^2<2$, we have
\begin{equation}
B_{(n)}= \frac{1}{a^2}\left[\mathcal{C}_1\cosh\left(\eta\sqrt{2-n^2}\right) +\mathcal{C}_2\sinh\left(\eta\sqrt{2-n^2}\right)\right]\,, \label{eqn:B-amplitude2}
\end{equation}
since $K=-1$.\footnote{The agreement between the sets (\ref{Aa0+1}), (\ref{eqn:B-amplitude1}) and (\ref{Aa-1}), (\ref{eqn:B-amplitude2}) becomes intuitively plausible once we recall that $B_a= \epsilon_{abc}{\rm D}^b\mathcal{A}^c$ to linear order on FRW backgrounds (see Eq.~(\ref{magn-field-pot}) in \S~\ref{ssVSPs} earlier).} This solution exhibits exponential behaviour closely analogous to that of the vector potential seen in (\ref{Aa-1}). More specifically, in open Friedmann models with conventional matter, the dominant magnetic mode of (\ref{eqn:B-amplitude2}) obeys the power law $B_{(n)}\propto a^{\sqrt{2-n^2}-2}$, as long as $0<n^2<2$~\cite{BT,T2}. Again, the reason for the qualitative change in the magnetic evolution is the negative curvature of the universe's spatial sections.

Following (\ref{amplAa}) and (\ref{eqn:B-amplitude2}), in Friedmann universes with hyperbolic spatial geometry, both the vector potential and the magnetic field are superadiabatically amplified, a term originally coined in gravitationally-wave studies~\cite{G}.\footnote{The reader is referred to~\cite{dGML} for a comparison of graviton production in closed and open Friedmann models.} It should be noted that, in our case, the superadiabatic amplification occurs despite the conformal invariance of the Maxwell field, which still holds. This happens because, in contrast to the flat FRW spacetime which is globally conformal to the Minkowski space, the conformal flatness of its curved counterparts is only local (e.g.~see~\cite{S}-\cite{IKM}). As a result, in the latter type of models, the electromagnetic wave equation acquires extra curvature-related terms and the familiar adiabatic decay law is not a priori guaranteed. Instead, on spatially open FRW backgrounds, the Maxwell field can be superadiabatically amplified (see also~\cite{BT} for further discussion).

\subsection{Gravitational-wave effects on electromagnetic 
signals}\label{ssG-WEEMSs}
Studies on the interaction between gravitational and electromagnetic waves have a long history, with most of the available treatments involving the electric and the magnetic fields directly (e.g.~see~\cite{Co}-\cite{BH}). In what follows, we will provide an alternative approach that involves the potentials of the Maxwell field.

\subsubsection{The Weyl-Maxwell coupling in Minkowski 
space}\label{sssW-MCMS}
Provided that the gravito-electromagnetic interaction takes place in the low-density interstellar space, away from massive compact stars, we may assume that the host environment is described by the Minkowski spacetime. There, we may also treat both the electromagnetic and the gravitational waves as test fields propagating in an otherwise empty and static space. In such an environment, the wave formulae of the vector and the scalar potentials (see Eqs.~(\ref{eqn:vec-pot}) and (\ref{eqn:scalar-pot}) in \S~\ref{ssWEPs}) linearise to
\begin{equation}
\ddot{\mathcal{A}}_a- {\rm D}^2\mathcal{A}_a= 0 \hspace{15mm} {\rm and} \hspace{15mm} \ddot{\Phi}- {\rm D}^2\Phi= 0\,,  \label{Mwp}
\end{equation}
respectively. The above accept simple plane-wave solutions of the form
\begin{equation}
\mathcal{A}_{(n)}= \mathcal{C}\sin(nt+\mathcal{\theta}_\mathcal{C}) \hspace{15mm} {\rm and} \hspace{15mm} \Phi_{(n)}= \mathcal{D}\sin(nt+\mathcal{\theta}_\mathcal{D})\,, \label{eqn:orig-scalar-pot}
\end{equation}
with $\mathcal{A}_{(n)}$ representing the $n$-th harmonic modes of the vector potential and $\Phi_{(n)}$ the one of its scalar counterpart.\footnote{Solution (\ref{eqn:orig-scalar-pot}a) follows after introducing the harmonic splitting $\mathcal{A}_a= \sum_n\mathcal{A}_{(n)}\mathcal{Q}_a^{(n)}$ (see also footnote~9), while in (\ref{eqn:orig-scalar-pot}b) we have assumed that $\Phi=\sum_n\Phi_{(n)}\mathcal{Q}^{(n)}$. In the latter case, $\mathcal{Q}^{(n)}$ are standard scalar harmonic functions, with ${\rm D}^2\mathcal{Q}^{(n)}=-n^2\mathcal{Q}^{(n)}$. Also, $\dot{\mathcal{Q}}^{(n)}=0={\rm D}_a\Phi_{(n)}$ by construction.} Given the flatness of the Minkowski background, $n$ is the physical wavenumber of the mode, with $n^2=n_an^a$ and $n_a$ representing the corresponding eigenvector. Also, $\mathcal{C}$, $\mathcal{D}$ and $\mathcal{\theta}_{\mathcal{C}}$, $\mathcal{\theta}_{\mathcal{D}}$ are the associated amplitudes and phase constants, to be determined by the initial conditions.

Within the framework of the 1+3 covariant approach, gravitational radiation is described by the electric and the magnetic components of the Weyl field (see Appendix~\ref{AppA3}). Also, isolating linear gravitational waves requires imposing a number of constraints to guarantee that only the pure-tensor part of the free gravitational field is accounted for~\cite{TCM,EMM}. In practice, this means ensuring that ${\rm D}^bE_{ab}=0={\rm D}^bH_{ab}$ and that only the transverse component of these traceless tensors survives. Given the absence of matter and the static nature of the Minkowski space, this is achieved by demanding that $\omega_a=0=\dot{u}_a$ to first order. These translate into the following linear relations
\begin{equation}
\dot{\sigma}_{ab}= -E_{ab} \hspace{10mm} {\rm and} \hspace{10mm} H_{ab}= {\rm curl}\sigma_{ab}\,,  \label{MsGW}
\end{equation}
between the Weyl tensors and the shear (see Eqs.~(\ref{shear-prop}) and (\ref{kcon3}) in Appendix~\ref{AppA2}). Therefore, in our environment, the linear evolution of both $E_{ab}$ and $H_{ab}$ is determined by shear perturbations and more specifically by the transverse (i.e.~the pure tensor -- ${\rm D}^b\sigma_{ab}=0$) part of the shear. The latter satisfies the wave equation~\cite{T3,KT}
\begin{equation}
\ddot{\sigma}_{ab}- {\rm D}^2\sigma_{ab}=0\,,  \label{Mwsh}
\end{equation}
which in turn admits the solution
\begin{equation}
\sigma_{(k)}= \mathcal{G}\sin(kt+\mathcal{\theta}_\mathcal{G})\,. \label{eqn:orig-shear}
\end{equation}
In the above $k$ is the physical wavenumber of the mode, with $k^2=k_ak^a$ and $k_a$ representing the corresponding wavevector. Also, $\mathcal{G}$ is the amplitude of the gravitational wave and $\mathcal{\theta}_\mathcal{G}$ is the associated phase.\footnote{In deriving Eq.~(\ref{eqn:orig-shear}) we have employed the harmonic splitting $\sigma_{ab}= \sigma_{(k)}\mathcal{Q}^{(k)}_{ab}$, where $\dot{\mathcal{Q}}^{(k)}_{ab}=0={\rm D}_a\sigma_{(k)}$. Also, $\mathcal{Q}^{(k)}_{ab}$ are pure-tensor harmonics that satisfy the constraints $\mathcal{Q}_{ab}^{(k)}=\mathcal{Q}_{(ab)}^{(k)}$, ${\rm D}^b\mathcal{Q}_{ab}^{(k)}=0$ and ${\rm D}^2\mathcal{Q}_{ab}^{(k)}=-k^2\mathcal{Q}_{ab}^{(k)}$.}

Solutions (\ref{eqn:orig-scalar-pot}) describe linear electromagnetic waves propagating on a Minkowski background in terms of the associated potentials, while solution (\ref{eqn:orig-shear}) does the same for gravitational radiation in terms of the corresponding shear perturbations. The interaction between these two sources is monitored by the wave formulae
\begin{equation}
\ddot{\tilde{\mathcal{A}}}_a- {\rm D}^2\tilde{\mathcal{A}}_a= 2\sigma_{ab}{\rm D}^b\Phi \hspace{15mm} {\rm and} \hspace{15mm}
\ddot{\tilde{\Phi}}- {\rm D}^2\tilde{\Phi}= 2\sigma_{ab}{\rm D}^{a}\mathcal{A}^b\,,   \label{eqn:scalar-pot-coupl}
\end{equation}
at the second perturbative level (see Eqs.~\eqref{eqn:vec-pot} and~\eqref{eqn:scalar-pot} in \S~\ref{ssWEPs}). Note that $\mathcal{A}_{a}$ and $\Phi$ represent the (linear) potentials prior to the gravito-electromagnetic interaction, while their ``tilded'' counterparts (i.e.~$\tilde{\mathcal{A}}_{a}$ and $\tilde{\Phi}$) are the (second order) potentials that emerged from the interaction. Also, in deriving \eqref{eqn:scalar-pot-coupl}, we have taken into account that, on our Minkowski background, the Gauss-Codacci equation (see (\ref{GC}) in Appendix~\ref{AppA3}) linearises to $\mathcal{R}_{ab}=E_{ab}$. It is also worth noting that the wave formulae (\ref{eqn:scalar-pot-coupl}a) and (\ref{eqn:scalar-pot-coupl}b) account for the ``backreaction'' of the scalar potential upon the its vector counterpart and vice-versa. Including these effects allows us to extend the analysis of~\cite{T3,KT}, where the analogous backreaction of the electric upon the magnetic component (and vice-versa) was bypassed.

We proceed to analyse the coupling between the Weyl and the Maxwell fields, by harmonically decomposing the gravitationally induced potentials. In other words, we set
\begin{equation}
\tilde{\mathcal{A}}_a= \tilde{\mathcal{A}}_{(\ell)}\tilde{\mathcal{Q}}^{(\ell)}_a \hspace{10mm} {\rm and} \hspace{10mm} \tilde{\Phi}= \tilde{\Phi}_{(\ell)}\tilde{\mathcal{Q}}^{(\ell)}\,,  \label{hsplit2}
\end{equation}
where $\ell$ is the physical wavenumber of the induced modes (with $\ell^2=\ell_a\ell^a$ and $\ell_a$ being the associated wavevector).\footnote{The vector and scalar harmonics seen in  (\ref{hsplit2}) are $\tilde{\mathcal{Q}}^{(\ell)}_a= \mathcal{Q}^{(n)}\mathcal{Q}^{(k)}_{ab}n^b$ and $\tilde{\mathcal{Q}}^{(\ell)}= \mathcal{Q}^{(k)}_{ab}\mathcal{Q}^a_{(n)}n^b$ by construction, where $n_a$ is the wavevector of the potentials. Note that, since $\dot{\mathcal{Q}}^{(n)}=0=\dot{\mathcal{Q}}^{(k)}_{ab}$, it follows that $\dot{\tilde{\mathcal{Q}}}^{(\ell)}_a=0= \dot{\tilde{\mathcal{Q}}}^{(\ell)}$ as well. In addition, recalling that ${\rm D}^2\mathcal{Q}^{(n)}=-n^2\mathcal{Q}^{(n)}$ and that ${\rm D}^2\mathcal{Q}^{(k)}_{ab}=-k^2\mathcal{Q}^{(n)}_{ab}$, one can show that ${\rm D}^2\tilde{\mathcal{Q}}^{(\ell)}_a= -\ell^2\tilde{\mathcal{Q}}^{(\ell)}_a$ and that ${\rm D}^2\tilde{\mathcal{Q}}^{(\ell)}= -\ell^2\tilde{\mathcal{Q}}^{(\ell)}$, with $\ell$ satisfying conditions (\ref{l-k-n}).} These are related to the wavevectors and the wavenumbers of the initially interacting sources by the expressions
\begin{equation}
\ell_a= k_a+ n_a \hspace{10mm} {\rm and} \hspace{10mm} \ell^2= n^2+ k^2+ 2nk\cos\phi\,,  \label{l-k-n}
\end{equation}
with $0\leq\phi\leq\pi$ representing the interaction angle of the original linear waves.

\subsubsection{Weyl-Maxwell resonances}\label{sssWMRs}
Substituting decompositions (\ref{hsplit2}), together with those of the initially interacting electromagnetic and gravitational signals (see footnotes~9 and~10) back into the second-order formulae (\ref{eqn:scalar-pot-coupl}a) and (\ref{eqn:scalar-pot-coupl}b), the latter take the form
\begin{equation}
\ddot{\tilde{\mathcal{A}}}_{(\ell)}+ \ell^{2}\tilde{\mathcal{A}}_{(\ell)}= 2\sigma_{(k)}\Phi_{(n)} \hspace{15mm} {\rm and} \hspace{15mm} \ddot{\tilde{\Phi}}_{(\ell)}+ \ell^2\tilde{\Phi}_{(\ell)}= 2\sigma_{(k)}\mathcal{A}_{(n)}\,,  \label{hwtAtF}
\end{equation}
respectively.\footnote{The phase factor ${\rm e}^{\imath\pi/2}$ in the 3-gradient of the potential has been ``absorbed'' into the associated wavevector.} Employing the linear solutions (\ref{eqn:orig-scalar-pot}) and (\ref{eqn:orig-shear}), the first of the above differential equations recasts as
\begin{equation}
\ddot{\tilde{\mathcal{A}}}_{(\ell)}+ \ell^2\tilde{\mathcal{A}}_{(\ell)}= \mathcal{E}\left\{\cos[(k-n)t+\theta_{\mathcal{E}_1}] -\cos[(k+n)t+\theta_{\mathcal{E}_2}]\right\}\,,  \label{ddcA}
\end{equation}
while the latter reads
\begin{equation}
\ddot{\Phi}_{(\ell)}+ \ell^2\Phi_{(\ell)}= \mathcal{M}\left\{\cos[(k-n)t+\theta_{\mathcal{M}_1}] -\cos[(k+n)t+\theta_{\mathcal{M}_2}]\right\}\,.  \label{ddPhi}
\end{equation}
Here, $\mathcal{E}=\mathcal{G}\mathcal{D}$ and  $\mathcal{M}= \mathcal{G}\mathcal{C}$ are the amplitudes of the gravitationally induced potential waves, while $\theta_{\mathcal{E}_{1,2}}= \theta_\mathcal{G}\mp\theta_\mathcal{D}$ and $\theta_{\mathcal{M}_{1,2}}=\theta_{\mathcal{G}}\mp \theta_{\mathcal{C}}$, $\theta_{J_{2}}\equiv \theta_{F}+\theta_{D}$ are the associated phases (all fixed at the onset of the gravito-electromagnetic interaction). According to (\ref{ddcA}) and (\ref{ddPhi}), the induced electromagnetic signal is driven by the superposition of two waves, with effective wave numbers $m_{1,2}=k\mp n$. Solving Eqs.~(\ref{ddcA}) and (\ref{ddPhi}) leads to
\begin{equation}
\tilde{\mathcal{A}}_{(\ell)}= \mathfrak{D}\sin(\ell t+\vartheta)+ \mathcal{K}_1\cos[m_1t+\theta_{\mathcal{E}_1}]- \mathcal{K}_2\cos[m_2t+\theta_{\mathcal{E}_2}]  \label{cA}
\end{equation}
and
\begin{equation}
\Phi_{(\ell)}= \mathfrak{D}\sin(\ell t+\vartheta)+ \mathcal{L}_1\cos[m_1t+\theta_{\mathcal{M}_1}]- \mathcal{L}_2\cos[m_2t+\theta_{\mathcal{M}_2}]\,,  \label{Phi}
\end{equation}
respectively. Note that $\mathfrak{D}$, $\vartheta$, $\mathcal{K}_{1,2}$ and $\mathcal{L}_{1,2}$ are constants determined at the onset of the Weyl-Maxwell interactions, with the latter two given by
\begin{equation}
\mathcal{K}_{1,2}= \frac{\mathcal{E}}{\ell^2-m_{1,2}^2} \hspace{15mm} {\rm and} \hspace{15mm} \mathcal{L}_{1,2}= \frac{\mathcal{M}}{\ell^{2}-m_{1,2}^2}\,.  \label{cKcL}
\end{equation}
Accordingly, the gravito-electromagnetic coupling leads to resonances when $\ell\rightarrow m_{1,2}=k\mp n$. In particular, when $\ell\rightarrow m_1=k-n$, relation~\eqref{l-k-n} implies that the two original waves propagate in opposite directions (i.e. $\phi\rightarrow\pi$). When $\ell\rightarrow m_2=k+n$, on the other hand, the original waves propagate along the same direction (i.e. $\phi\rightarrow0$). Note that these results are in agreement with those obtained after employing the electromagnetic fields instead of their potentials~\cite{T3,KT}.

Despite the appearances, the resonances identified in this section do not generally suggest an arbitrarily strong enhancement of the emerging electromagnetic wave. Instead, and in analogy with forced harmonic oscillations in classical mechanics, the aforementioned resonances imply linear (in time) growth for the amplitude of the electromagnetic signal. Typically, this requires the ``smooth'' transition between the potentials prior and after the interaction, namely it follows naturally after imposing the conditions $\tilde{\mathcal{A}}_{(\ell)}=\mathcal{A}_{(n)}$ and $\tilde{\Phi}_{(\ell)}=\Phi_{(n)}$ at the onset of the Weyl-Maxwell coupling. We refer the reader to \S~III in~\cite{KT} for a thorough discussion of the gravito-electromagnetic case, as well as to~\cite{LL} for the presentation of the mechanical analogue.

\section{Discussion}\label{sD}
Electromagnetic fields appear everywhere in the universe, either in the form of ``individual'' electric and magnetic fields, or as traveling electromagnetic radiation. A special feature of the Maxwell field, which separates it from the other known energy sources, is its vector nature. The latter ensures a purely geometrical coupling between electromagnetism and spacetime curvature that is manifested through the Ricci identities and goes beyond the standard interplay between matter and geometry introduced by Einstein's equations. As a result, the evolution of electric and magnetic fields, as well as the propagation of electromagnetic signals, are affected by the curvature of the host spacetime via both of the aforementioned relations.

Most of the available studies employ, as well  as target, the electric and magnetic components directly. Here, we provide an alternative (fully general relativistic) treatment, which uses the 1+3 covariant formalism and involves the electromagnetic vector and scalar potentials. Although the latter may not be directly measurable physical entities, their existence is theoretically allowed by the form of Maxwell's equations. In addition, the temporal and spatial gradients of the vector and scalar potentials give rise to the actual electric and magnetic fields. Therefore, depending on the nature of the problem in hand, one is in principle free to chose either description when analysing electromagnetic phenomena. Given that an 1+3 covariant treatment of electromagnetic fields in curved spacetimes was already given in~\cite{T1}, we have provided here a supplementary study involving the scalar and the vector potentials of the Maxwell field.

We began by introducing a family of observers, which facilitated the 1+3 splitting of the spacetime into a temporal direction (along the observers' 4-velocity vector) and 3-dimensional spatial hypersurfaces orthogonal to it. This in turn allowed us to decompose the electromagnetic 4-potential into its timelike and spacelike parts, respectively represented by the associated scalar and vector potentials. The latter were shown to satisfy wave-like equations, which were directly derived from Maxwell's formulae and contained driving terms reflecting the nature and the material content of the host spacetime. Given that the electromagnetic potential trivially satisfies one of Maxwell's equations, both of the aforementioned wave formulae were derived from the other. More specifically, Faraday's law leads to the wave equation of the vector potential and Coulomb's law to that of its scalar counterpart. In the case of the vector potential, some of the aforementioned driving terms were due to the nonzero spacetime curvature. We found, in particular, that both the spatial and the Weyl parts of the curvature can affect the evolution of the vector potential, through the latter's purely geometrical coupling to the spacetime geometry (mediated by the Ricci identities). No such coupling holds for the scalar potential, which explains why there were no direct spacetime curvature effects in the wave equation of the latter.

Since our principal aim was to study the Maxwell field in curved spacetimes, we applied the wave formula of the vector potential to a Friedmann model with non-Euclidean spatial geometry. Confining to the linear regime of an almost-FRW universe, we found that in spatially closed models the potential oscillates with an amplitude that decays inversely proportional to the cosmological scale factor, just like it does in spatially flat Friedmann models. The only effect of the positive curvature, was to increase the frequency of the oscillation. On the other hand, the hyperbolic spatial geometry of the open FRW universes modified the evolution of the vector potential in a more ``dramatic'' way. There, the model's negative curvature changed the standard oscillatory behaviour to a power-law evolution. Not surprisingly, this qualitative change was found to occur on sufficiently large scales, where the effects of the non-Euclidean geometry become more prominent. Exactly analogous curvature effects were also observed during the evolution of source-free electric and magnetic fields in perturbed Friedmann models~\cite{BT,T2}.

We then turned to astrophysical environments and employed the electromagnetic potentials to investigate the coupling between the Maxwell and the Weyl fields in the low-density interstellar space. In practice, this meant using the potentials to study the interaction between propagating gravitational and electromagnetic waves on a Minkowski background at the second perturbative order. Given that gravity-wave (i.e.~pure tensor) perturbations are monitored by shear distortions, we included the driving effects of the latter into the wave formulae of the scalar and the vector potentials. Our results showed that the gravitationally induced electromagnetic potentials perform forced oscillations, driven by the coupling between the originally interacting waves. This immediately opens the possibility of resonances, which in our case occur when the initial electromagnetic and gravitational waves propagate along the same, or in the opposite, direction. In most realistic situations, the aforementioned gravito-electromagnetic resonances lead to the linear amplification of the emerging electromagnetic signal. Exactly analogous resonances and amplification effects were reported in the studies of~\cite{T3,KT}, which employed the electric and the magnetic components of the Maxwell field, instead of the potentials. We finally note that, in the present analysis, we also accounted for the backreaction effects between the scalar and the vector potentials, while those of~\cite{T3,KT} bypassed similar backreaction effects between the electric and the magnetic fields. This underlines the considerable technical simplification that one can achieve by involving the electromagnetic potentials instead of the actual electric and magnetic fields.

The compete agreement between our results and those of the previous more conventional studies, together with the technical advantages the use of the potentials seems to bring in, suggest that the formalism introduced and developed here could prove particularly useful when probing the behaviour of electromagnetism in technically demanding astrophysical and cosmological environments. Here, we considered the highly symmetric Minkowski and FRW backgrounds. In principle, however, our analysis can be also applied to, say, the vicinity and perhaps the interior of massive compact stars, or to the very early stages of the universe's evolution and to the study of the Cosmic Microwave Background (CMB).

\appendix

\section{1+3 covariant formalism}\label{AppA}
In what follows, we outline the basic principles of the 1+3 covariant formalism and also provide the key relations used in this study. More details, together with further discussion and references, the reader can find in the extensive reviews of~\cite{TCM,EMM}.

\subsection{Spacetime splitting}\label{AppA1}
In the context of the 1+3 covariant approach to general relativity, the 4-dimensional spacetime decomposes into a temporal direction and a 3-dimensional space orthogonal to it~\cite{TCM,EMM}. This splitting is achieved by introducing a family of (fundamental) observers, moving along their timelike worldlines. These have parametric equations of the form $x^{a}=x^{a}(\tau)$, where $\tau$ is the observer's proper time.\footnote{Throughout this study, Latin indices vary between 0 and 3 and we have set the velocity of light to unity.} The tangent vector to these worldlines is the observer's 4-velocity (with $u^{a}={\rm d}x^{a}/{\rm d}\tau$ and $u^au_a=-1$) and defines their temporal direction. Then, assuming that $g_{ab}$ is the spacetime metric, the symmetric tensor $h_{ab}=g_{ab}+u_au_b$, with $h_{ab}u^{b}=0$, $h_a{}^a=3$, $h_a{}^ch_{cb}=h_{ab}$ by construction, projects orthogonal to the $u_a$-field and into the observer's 3-D rest-space.

On using the $u_a$-field and the associated projection tensor $h_{ab}$, one can decompose every spacetime vector and tensor, every operator and every equation into their temporal and spatial components. For instance, the 4-vector $V_a$ decomposes as
\begin{equation}
V_a= \mathcal{V}u_a+ \mathcal{V}_a\,,  \label{Vsplit}
\end{equation}
where $\mathcal{V}=-V_au^a$ is the timelike part parallel to $u_a$ and $\mathcal{V}_a=h_a{}^bV_a$ is its spacelike counterpart orthogonal to $u_a$. Similarly, the symmetric second-rank tensor $T_{ab}$ splits as
\begin{equation}
T_{ab}= tu_au_b+ \frac{1}{3}\left(T+t\right)h_{ab}+ 2u_{(a}t_{b)}+ t_{ab}\,,  \label{Tsplit}
\end{equation}
with $T=T_a{}^a$, $t=T_{ab}u^au^b$, $t_a=-h_a{}^bT_{bc}u^c$ and $t_{ab}=h_{\langle a}{}^ch_{b\rangle}{}^dT_{cd}$.\footnote{Recall that round brackets denote symmetrisation, square ones antisymmetrisation and angular brackets describe the symmetric traceless part of orthogonally projected second-rank tensors (e.g.~$T_{\langle ab\rangle}=T_{(ab)}-(1/3)T_c{}^ch_{ab}$).} The above decomposition follows from the expression $T_{ab}=g_{ac}g_{bd}T^{cd}= (h_{ac}-u_au_c)(h_{bd}-u_bu_d)T^{cd}$ and its most familiar application is on the energy-momentum tensor of a general imperfect fluid (e.g.~see~\cite{TCM,EMM}). An additional useful splitting is that of the 4-D Levi-Civita tensor ($\eta_{abcd}=\eta_{[abcd]}$). Relative to the $u_a$-field, the latter decomposes according to
\begin{equation}
\eta_{abcd}= 2u_{[a}\epsilon_{b]cd}- 2\epsilon_{ab[c}u_{d]}
\end{equation}
where $\epsilon_{abc}=\epsilon_{[abc]}=\eta_{abcd}u^d$ is the Levi-Civita tensor of the 3-D spatial hypersurfaces. Then, $\epsilon_{abc}u^c=0$ and  $\epsilon_{abc}\epsilon^{def}= 3!h_{[a}{}^dh_b{}^eh_{c]}^f$ by construction.

Once the time-direction and the orthogonal 3-space have been introduced, one needs to define temporal and spatial differentiation. For a general tensor field $T_{ab\cdots}{}^{cd\cdots}$, the time and the 3-space derivatives are respectively given by
\begin{equation}
\dot{T}_{ab\cdots}{}^{cd\cdots}= u^{e}\nabla_{e}T_{ab\cdots}{}^{cd\cdots} \hspace{10mm} {\rm and} \hspace{10mm} {\rm D}_{e}T_{ab\cdots}{}^{cd\cdots}= h_e{}^sh_a{}^fh_b{}^ph_q{}^ch_r{}^d\cdots \nabla_{s}T_{fp\cdots}{}^{qr\cdots}\,,  \label{tempspatder}
\end{equation}
with $\nabla_a$ representing the 4-D covariant derivative operator. It follows that ${\rm D}_ah_{bc}=0={\rm D}_d\epsilon_{abc}$ and that $\dot{\epsilon}_{abc}=3u_{[a}\epsilon_{bc]d}\dot{u}^d$, with $\dot{u}_a$ being the 4-acceleration (see Appendix~\ref{AppA2} next).

\subsection{Covariant kinematics}\label{AppA2}
All the information regarding the kinematic evolution of the 4-velocity field is encoded in its covariant gradient. The latter decomposes into the irreducible kinematic variables of the motion according to
\begin{equation}
\nabla_bu_a= \sigma_{ab}+ \omega_{ab}+ \frac{1}{3}\,\Theta h_{ab}- \dot{u}_au_b\,,
\end{equation}
with $\sigma_{ab}\equiv {\rm D}_{\langle b}u_{a\rangle}$, $\omega_{ab}\equiv {\rm D}_{[b}u_{a]}$, $\Theta\equiv \nabla^{a}u_{a}={\rm D}^{a}u_{a}$ and $\dot{u}_{a}\equiv u^{b}\nabla_{b}u_{a}$ respectively representing the shear and the vorticity tensors, the volume expansion/contraction scalar and the 4-acceleration vector. The shear monitors distortion in the shape of a moving fluid element and nonzero vorticity implies rotation. The volume scalar, on the other hand, determines the expansion/contraction of the fluid (when it is positive/negative). Finally, a nonzero 4-acceleration reveals the presence of non-gravitational forces, which in turn ensures non-geodesic motion. Note that the vorticity vector leads to the vector $\omega_a=\epsilon_{abc}\omega^{bc}/2$, which determines the rotation axis. Also, the volume scalar is typically used to define a representative length-scale ($a$), so that $\dot{a}/a=\Theta/3$. In cosmological studies, $a$ is identified with the scale factor of the universe, which the volume scalar and the Hubble parameter ($H$) are related by $\Theta/3=H$.

The evolution of the volume scalar, the shear and the vorticity is monitored by a set of three propagation equations, supplemented by an equal number of constraints. These are obtained after applying the Ricci identity (see (\ref{eqn:Ricc-Electr}) in \S~\ref{ssEMFC}) to the 4-velocity field. More specifically, the timelike component of the resulting expression leads to the Raychaudhuri equation
\begin{equation}
\dot{\Theta}= -\frac{1}{3}\,\Theta^2- \frac{1}{2}\,\kappa(\rho+3p)- 2\left(\sigma^2-\omega^2\right)+ {\rm D}^a\dot{u}_a+ \dot{u}^a\dot{u}_a\,,  \label{Ray}
\end{equation}
to the shear evolution formula
\begin{equation}
\dot{\sigma}_{\langle ab\rangle}= -\frac{2}{3}\,\Theta\sigma_{ab}- \sigma_{c\langle a}\sigma^c{}_{b\rangle}- \omega_{\langle a}\omega_{b\rangle}+ {\rm D}_{\langle a}\dot{u}_{b\rangle}+ \dot{u}_{\langle a}\dot{u}_{b\rangle}- E_{ab}+ \frac{1}{2}\,\kappa\pi_{ab}  \label{shear-prop}
\end{equation}
and to the propagation equation of the vorticity tensor
\begin{equation}
\dot{\omega}_{\langle ab\rangle}= -{2\over3}\,\Theta\omega_{ab}+ {\rm D}_{[b}\dot{u}_{a]}- 2\sigma_{c[a}\omega^c{}_{b]}\,,  \label{dotvorten}
\end{equation}
where $\dot{\sigma}_{\langle ab\rangle}= h_a{}^ch_b{}^d\dot{\sigma}_{cd}$ and $\dot{\omega}_{\langle ab\rangle}= h_a{}^ch_b{}^d\dot{\omega}_{cd}$ by construction. Note that, recalling the $\omega_{ab}=\epsilon_{abc}\omega^c$, one could replace the above with the evolution formula of the vorticity vector
\begin{equation}
\dot{\omega}_{\langle a\rangle}= -\frac{2}{3}\,\Theta\omega_a- \frac{1}{2}\,{\rm curl}\dot{u}_a+\sigma_{ab}\omega^b\,.
\end{equation}

The propagation formulae of the irreducible kinematic variables are supplemented by three constraints. These are obtained from the spatial part of the aforementioned Ricci identities and they are given by
\begin{equation}
{\rm D}^b\sigma_{ab}= {2\over3}\,{\rm D}_a\Theta+ {\rm curl}\omega_a+ 2\epsilon_{abc}\dot{u}^b\omega^c- \kappa q_a\,, \hspace{15mm} {\rm D}^a\omega_a= \dot{u}^a\omega_a  \label{kcon12}
\end{equation}
and
\begin{equation}
H_{ab}= {\rm curl}\sigma_{ab}+ {\rm D}_{\langle a}\omega_{b\rangle}+ 2\dot{u}_{\langle a}\omega_{b\rangle}\,.  \label{kcon3}
\end{equation}
Note that $\sigma^2=\sigma_{ab}\sigma^{ab}/2$ and $\omega^2=\omega_{ab}\omega^{ab}/2=\omega_a\omega^a$ are the (square) magnitudes of the shear and the vorticity respectively, while $E_{ab}$ and $H_{ab}$ are the electric and the magnetic components of the Weyl tensor (see Appendix~\ref{AppA3} next). Finally, ${\rm curl}\sigma_{ab}=\epsilon_{cd\langle a}{\rm D}^c\sigma_{b\rangle}{}^d$ by construction.

\subsection{Spatial and Weyl curvature}\label{AppA3}
The Riemann tensor ($R_{abcd}$), which determines the curvature of the 4-D spacetime satisfies the symmetries $R_{abcd}=R_{cdab}$, $R_{abcd}=R_{[ab][cd]}$ and $R_{a[bcd]}=0$. Also, the trace of $R_{abcd}$ leads to the symmetric Ricci tensor via the contraction $R_{ab}=R^c{}_{acb}$. The latter, together with the Ricci scalar $R=R^a{}_a$, determine the local gravitational field due to the presence of matter by means of Einstein's equations (see expression (\ref{EFE}) in \S~\ref{ssEMFC}).

The (intrinsic) curvature of the 3-D hypersurfaces orthogonal to the the observers' 4-velocity is determined by the associated 3-Riemann tensor (see Eqs.~(\ref{3-D-Ricci-EL}) in~\S~\ref{ssEMFC}), given by
\begin{equation}
\mathcal{R}_{abcd}= h_a{}^eh_b{}^fh_c{}^qh_d{}^sR_{efqs}- {\rm D}_cu_a{\rm D}_du_b+ {\rm D}_du_a{\rm D}_cu_b\,,  \label{3Riemann}
\end{equation}
with the 4-velocity gradient ${\rm D}_bu_a=(\Theta/3)h_{ab}+ \sigma_{ab}+\omega_{ab}$ describing the extrinsic curvature. When there is no vorticity (i.e.~for $\omega_{ab}=0$), the 3-Riemann tensor share all the symmetries of its 4-D counterpart. In the opposite case, we have $\mathcal{R}_{abcd}=\mathcal{R}_{[ab][cd]}$ only. Then, the corresponding 3-Ricci tensor $\mathcal{R}_{ab}= \mathcal{R}^c{}_{acb}$ satisfies the Gauss-Codacci equation
\begin{eqnarray}
\mathcal{R}_{ab}&=& {2\over3}\left(\kappa\rho-{1\over3}\,\Theta^2+\sigma^2-\omega^2\right)h_{ab}- E_{ab}+ {1\over2}\,\kappa\pi_{ab}- {1\over3}\,\Theta(\sigma_{ab}+\omega_{ab}) \nonumber\\ &&+\sigma_{c\langle a}\sigma^c{}_{b\rangle}- \omega_{c\langle a}\omega^c{}_{b\rangle}+ 2\sigma_{c[a}\omega^c{}_{b]}\,.  \label{GC}
\end{eqnarray}
It follows that, in contrast to its 4-D counterpart, $\mathcal{R}_{ab}$ is no longer symmetric. Instead, in rotating spacetimes, the 3-Ricci tensor has an antisymmetric part that is given by
\begin{equation}
\mathcal{R}_{[ab]}= -{1\over3}\,\Theta\omega_{ab}+ 2\sigma_{c[a}\omega^c{}_{b]}\,.  \label{cR[ab]}
\end{equation}
Finally, the trace of (\ref{GC}) leads to the 3-Ricci scalar $\mathcal{R}=\mathcal{R}^a{}_a=2[\rho-(\Theta^2/3)+\sigma^2 -\omega^2]$, which measures the mean curvature of the 3-D spatial sections.

The long-range gravitational field, namely tidal forces and gravity waves are monitored by the Weyl curvature tensor ($\mathcal{C}_{abcd}$), which satisfies the relation
\begin{equation}
\mathcal{C}_{abcd}= R_{abcd}- {1\over2}\left(g_{ac}R_{bd}+g_{bd}R_{ac} -g_{bc}R_{ad}-g_{ad}R_{bc}\right)+ {1\over6}\,R\left(g_{ac}g_{bd}-g_{ad}g_{bc}\right)\,,  \label{Weyl}
\end{equation}
share all the symmetries of the Riemann tensor and it is also trace-free. In addition, relative to the $u_a$-field, the Weyl tensor splits into an electric and a magnetic component given by
\begin{equation}
E_{ab}= \mathcal{C}_{acbd}u^cu^d \hspace{10mm} {\rm and} \hspace{10mm}
H_{ab}= {1\over2}\,\epsilon_a{}^{cd}\mathcal{C}_{cdbe}u^e\,,  \label{EMWeyl}
\end{equation}
both of which are symmetric, traceless and ``live'' in the observers rest-space (i.e.~$E_{ab}u^b=0=H_{ab}u^b$). Employing the above, the Weyl curvature tensor decomposes as
\begin{equation}
\mathcal{C}_{ab}{}^{cd}= 4\left(u_{[a}u^{[c}+h_{[a}{}^{[c}\right)E_{b]}{}^{d]}+
2\epsilon_{abe}u^{[c}H^{d]e}+ 2u_{[a}H_{b]e}\epsilon^{cde}\,,  \label{Weylsplit}
\end{equation}
relative to the $u_a$-field. Note that the electric and the magnetic Weyl tensors are monitored by a set of two propagation and two constraint equations, which share a number of common features with Maxwell's formulae. This resemblance, which has been thought as a possible sign of a closer underlying connection between the electromagnetic and the gravitational fields, has been the subject of debate for many decades.

\section{Deriving the wave equations for the potentials}\label{AppB}
This part of the Appendix provides guidance and some of the key steps leading to the wave formulae of the vector and the scalar electromagnetic potentials given in \S~\ref{ssWEPs}.

\subsection{The wave formula for the vector potential}\label{AppB1}
The wave formula for the vector potential (see Eq.~(\ref{eqn:vec-pot}) in \S~\ref{ssWEPs}) follows after combining expressions (\ref{el-field-pot}) and (\ref{magn-field-pot}) with Ampere's law (\ref{eqn:el-field-prop}). In particular, taking the time derivative of (\ref{el-field-pot}), using Raychaudhuri's equation (\ref{Ray}), together with the propagation formulae of the shear and the vorticity tensors (see (\ref{shear-prop}) and (\ref{dotvorten}) respectively), one arrives at
\begin{eqnarray}
\dot{E}_{\langle a\rangle}&=& -\ddot{\mathcal{A}}_{\langle a\rangle}- {1\over3}\,\Theta\dot{\mathcal{A}}_{\langle a\rangle}- (\sigma_{ab}-\omega_{ab})\dot{\mathcal{A}}^b+ {1\over3}\left[{1\over3}\,\Theta^2+{1\over2}\,\kappa(\rho+3p) +2\left(\sigma^2-\omega^2\right)-{\rm D}_b\dot{u}^b\right] \mathcal{A}^b \nonumber\\ &&+\left[{2\over3}\,\Theta(\sigma_{ab}-\omega_{ab}) +\sigma_{c\langle a}\sigma^c{}_{b\rangle}+\omega_{\langle a}\omega_{b\rangle}-2\sigma_{c[a}\omega^c{}_{b]}- {1\over2}\,\kappa\pi_{ab}- {\rm D}_{\langle b}\dot{u}_{a\rangle} +{\rm D}_{[b}\dot{u}_{a]}\right]\mathcal{A}^b \nonumber\\ &&+E_{ab}\mathcal{A}^b- 2\dot{\Phi}\dot{u}_a- \Phi\ddot{u}_{\langle a\rangle}- {\rm D}_a\dot{\Phi}+ {1\over3}\,\Theta{\rm D}_a\Phi+ (\sigma_{ab}-\omega_{ab}){\rm D}^b\Phi\,.  \label{AppB11}
\end{eqnarray}
Note that in deriving the above, which expresses the left-hand side of Eq.~(\ref{eqn:vec-pot}) in terms of the electromagnetic vector and scalar potentials, we have also used the auxiliary relation
\begin{equation}
h_a{}^b\left({\rm D}_a\Phi\right)^{\cdot}= \dot{\Phi}\dot{u}_a+ {\rm D}_a\dot{\Phi}- {1\over3}\,\Theta{\rm D}_a\Phi- (\sigma_{ab}-\omega_{ab}){\rm D}^b\Phi\,,  \label{AppB12}
\end{equation}
monitoring the commutation between the spatial and the temporal derivatives of $\Phi$.

The terms on the right-hand side of Ampere's law are also expressed in terms of the aforementioned potentials by means of (\ref{el-field-pot}) and (\ref{magn-field-pot}). The most involved derivation is that of ${\rm curl}B_a$, since it requires the use of the 3-Ricci identities (see expression (\ref{3-D-Ricci-EL}) in \S~\ref{ssEMFC}). In so doing and after applying the Lorenz-gauge condition (see Eq.~(\ref{Lorenz-gauge}) in \S~\ref{ssVSPs}) twice, we obtain
\begin{eqnarray}
{\rm curl}B_a&=& -{\rm D}^2\mathcal{A}_a+ \mathcal{R}_{ba}\mathcal{A}^b+ 2\omega_{ab}\dot{\mathcal{A}}^b- {1\over3}\left({\rm D}_b\dot{u}^b -{1\over3}\,\dot{u}_b\dot{u}^b\right)\mathcal{A}_a+ {1\over3}\,\dot{u}_{\langle a}\dot{u}_{b\rangle}\mathcal{A}^b \nonumber\\ &&-\mathcal{A}^b\left({\rm D}_{\langle b}\dot{u}_{a\rangle}-{\rm D}_{[b}\dot{u}_{a]}\right)- \dot{u}^b\left({\rm D}_{\langle b}\mathcal{A}_{a\rangle}-{\rm D}_{[b}\mathcal{A}_{a]}\right)+ {1\over3}\,\dot{\Phi}\dot{u}_a \nonumber\\ &&+\left({1\over3}\,\Theta\dot{u}_a-{\rm D}_a\Theta- 2{\rm curl}\omega_a\right)\Phi-{\rm D}_a\dot{\Phi}- \Theta{\rm D}_a\Phi- 2\omega_{ab}{\rm D}^b\Phi\,,  \label{AppB13}
\end{eqnarray}
where $\mathcal{R}_{ab}$ satisfies the Gauss-Codacci equation (see expression (\ref{GC}) in Appendix~\ref{AppA3}). Using the auxiliary relations given above and following the recommended steps, one may recast Ampere's law into the wave-like formula (\ref{eqn:vec-pot}), governing the evolution of the vector potential in an arbitrary Riemannian spacetime.

\subsection{The wave formula for the scalar potential}\label{AppB2}
The wave formula for the scalar potential (see Eq.~(\ref{eqn:scalar-pot}) in \S~\ref{ssWEPs}) is obtained after substituting expressions (\ref{el-field-pot}) and (\ref{magn-field-pot}), into Coulomb's law (see (\ref{magn-div}a) in \S~\ref{ssMEs}). To begin with, taking the spatial divergence of (\ref{el-field-pot}), we initially obtain
\begin{eqnarray}
{\rm D}^aE_a&=& {\rm D}^a\dot{\mathcal{A}}_{\langle a\rangle}- {\rm D}^2\Phi- \dot{u}_a{\rm D}^a\Phi- \Phi{\rm D}^a\dot{u}_a- {1\over3}\mathcal{A}^a{\rm D}_a\Theta- \mathcal{A}^b{\rm D}^a\left(\sigma_{ba}+\omega_{ba}\right) \nonumber\\ && -{1\over3}\,\Theta{\rm D}_a\mathcal{A}^a- \left(\sigma_{ba}+\omega_{ba}\right){\rm D}^a\mathcal{A}^b\,.  \label{AppB21}
\end{eqnarray}
Employing the Ricci identities (see Eq.~(\ref{eqn:Ricci-id-4pot}) in \S~\ref{ssEMPC}) and using the symmetries of the Riemann tensor (see \S~\ref{AppA3} previously), the first term on the right-hand side of the above reads
\begin{eqnarray}
{\rm D}^a\dot{\mathcal{A}}_{\langle a\rangle}&=& {1\over3}\,\Theta{\rm D}^a\mathcal{A}_a+ \left({\rm D}^a\mathcal{A}_a\right)^{\cdot}- {2\over3}\,\Theta\dot{u}_a\mathcal{A}^a- \dot{u}_a\dot{\mathcal{A}}^a+ R_{ab}u^a\mathcal{A}^b+ \left(\sigma_{ab}+\omega_{ab}\right)\mathcal{A}^a\dot{u}^b \nonumber\\ &&+\left(\sigma_{ba}+\omega_{ba}\right){\rm D}^a\mathcal{A}^b\,,  \label{AppB22}
\end{eqnarray}
with $R_{ab}=R^c{}_{acb}$ representing the 4-D Ricci tensor. Substituting this result back into Eq.~(\ref{AppB21}), adopting the Lorenz-gauge (i.e.~imposing condition (\ref{Lorenz-gauge}) in \S~\ref{ssVSPs}), employing Raychaudhuri's formula (see Eq.~(\ref{Ray}) in Appendix~\ref{AppA2}), using constraint (\ref{kcon12}a), while also taking into account that $R_{ab}u^a\mathcal{A}^b= T_{ab}u^a\mathcal{A}^b=-\kappa q_a\mathcal{A}^a$ (see footnote~4 in \S~\ref{ssEMVs}) and keeping in mind that $\omega_{abc}=\epsilon_{abc}\omega^c$ (see Appendix~\ref{AppA2}), we arrive at
\begin{eqnarray}
{\rm D}^aE_a&=& \ddot{\Phi}- {\rm D}^2\Phi+ {5\over3}\,\Theta\dot{\Phi}- \left[{1\over2}\,\kappa(\rho+3p)- {1\over3}\,\Theta^2+2\left(\sigma^2-\omega^2\right) -\dot{u}_a\dot{u}^a\right]\Phi- \dot{u}^a{\rm D}_a\Phi \nonumber\\ &&-\left[{\rm D}_a\Theta-{4\over3}\,\Theta\dot{u}_a+2{\rm curl}\omega_a-2\kappa q_a+(\sigma_{ab}+3\omega_{ab})\dot{u}^b -\ddot{u}_a\right]\mathcal{A}^a+ 2\dot{u}_a\dot{\mathcal{A}}^a \nonumber\\ &&-2\sigma_{ab}{\rm D}^b\mathcal{A}^a\,.  \label{AppB23}
\end{eqnarray}
At the same time, the right-hand side of expression (\ref{magn-field-pot}) gives $2\omega^aB_a=2\omega^a{\rm curl}\mathcal{A}_a-4\omega^2\Phi$. Combining the latter with (\ref{AppB23}), we can finally recast Eq.~(\ref{magn-div}a) into the wave-formula (\ref{eqn:scalar-pot}) of the scalar potential.\\

\textbf{Acknowledgements:} The authors were supported by the Hellenic Foundation for Research and Innovation (H.F.R.I.), under the ‘First Call for H.F.R.I. Research Projects to support Faculty members and Researchers and the procurement of high-cost research equipment Grant’ (Project Number: 789). PM also acknowledges
support from the Foundation for European Education and Culture.


\begin{thebibliography}{99}
\bibitem{LMEN} R. De Luca, M. di Mauro, S. Esposito and A. Naddeo, \textit{Feynman's different approach to electromagnetism}, Eur. J. Phys. \textbf{40}, 065205 (2019).
\bibitem{HH} J.A. Heras and R. Heras, \textit{On Feynman's handwritten notes on electromagnetism and the idea of introducing potentials before fields}, Eur. J. Phys. \textbf{41}, 035202 (2020).
\bibitem{DWB} B.S. DeWitt and R.W. Brehme, \textit{Radiation damping in a gravitational field}, Ann. Phys. (N.Y.) \textbf{9}, 220 (1960).
\bibitem{H} J. Hobbs, \textit{ A vierbein formalism of radiaton damping}, Ann. Phys. (N.Y.) \textbf{47}, 141 (1968).
\bibitem{QW} T.C. Quinn and R.M. Wald, \textit{An axiomatic approach to electromagnetic and gravitational radiation reaction of particles in curved space-time}, Phys. Rev. D \textbf{56}, 3381 (1997).
\bibitem{BT} J.D. Barrow and C.G. Tsagas, \textit{Slow decay of magnetic fields in open Friedmann universes}, Phys. Rev. D \textbf{77}, 107302 (2008).
\bibitem{AH} F.A. Asenjo and S.A. Hojman, \textit{Do electromagnetic waves always propagate along null geodesics?}, Class. Quantum Grav. \textbf{34}, 205011 (2017).
\bibitem{CPS} Y.-Z. Chu, K. Pasmatsiou and G.D. Starkman, \textit{Finite-size effects on the self-force}, Phys. Rev. D \textbf{101}, 104020 (2020).
\bibitem{PM} D. Pugliese and G. Montani, \textit{Aspects of GRMHD in high-energy astrophysics: geometrically thick disks and tori agglomerates around spinning black holes}, Gen. Rel. Grav. \textbf{53}, 51 (2021).
\bibitem{MTW} C.W. Misner, K.S. Thorne and J.A. Wheeler, \textit{Gravitation} (Freeman, San Francisco, 1973).
\bibitem{W} R.M. Wald, \textit{General Relativity} (University of Chicago Press, Chicago, 1984).
\bibitem{N} T.W. Noonan, \textit{Huygens's principle for the electromagnetic vector potential in Riemannian spacetimes}, Astrophys. J. \textbf{341}, 786 (1989).
\bibitem{DFC} F. De Felice and C. Clark, \textit{Relativity in Curved Manifolds} (Cambridge University Press, Cambridge, 1990).
\bibitem{T1} C.G. Tsagas,  \textit{Electromagnetic fields in curved spacetimes}, Class. Quantum Grav. \textbf{22}, 393 (2005).
\bibitem{TCM} C.G. Tsagas, A. Challinor and R. Maartens, \textit{Relativistic cosmology and large-scale structure}, Phys. Rep. \textbf{465}, 61 (2008).
\bibitem{EMM} G.F.R. Ellis, R. Maartens and M.A.H. MacCallum, \textit{Relativistic Cosmology} (Cambridge University Press, Cambridge, 2012).
\bibitem{T3} C.G. Tsagas, \textit{Gravitoelectromagnetic resonances}, Phys. Rev. D \textbf{84} (2011).
\bibitem{KT} A.P. Kouretsis and C.G. Tsagas, \textit{Gravito-electromagnetic resonances in Minkowski space}, Phys. Rev. D \textbf{88}, 044006 (2013).
\bibitem{TMcD} K.S. Thorne and D. MacDonald, \textit{Electrodynamics in Curved Spacetime - 3+1 Formulation}, MNRAS \textbf{198}, 339 (1982).
\bibitem{J} J.D. Jackson, \textit{Classical Electrodynamics} (Whiley, New Jersey, 1998).
\bibitem{EBH} G.F.R. Ellis, M. Bruni and J. Hwang, \textit{Density-gradient-vorticity relation in perfect-fluid Robertson-Walker perturbations}, Phys. Rev. D \textbf{42}, 1035 (1990).
\bibitem{LKNS} V.A. De Lorenci, R. Klippert, M. Novello and J.M. Salim, \textit{Nonlinear electrodynamics and FRW cosmology}, Phys. Rev. D \textbf{65}, 063501 (2002)
\bibitem{K} A.P. Kouretsis, \textit{Cosmic magnetization in curved and Lorentz violating space–times}, Eur. Phys. J. C \textbf{74}, 2879 (2014).
\bibitem{T2} C.G. Tsagas, \textit{On the magnetic evolution in Friedmann universes and the question of cosmic magnetogenesis}, Symmetry \textbf{8}, 122 (2016).
\bibitem{G} L.P. Grishchuk, \textit{Amplification of gravitational waves in an isotropic universe}, Sov. Phys. JETP \textbf{40}, 409 (1974).
\bibitem{dGML} M.R. de Garcia Maia, J.A.S. Lima, \textit{Graviton production in elliptical and hyperbolic universes}, Phys. Rev. D \textbf{54}, 6111 (1996).
\bibitem{S} H. Stefani, \textit{Introduction to General Relativity} (Cambridge University Press, Cambridge, 1990).
\bibitem{KB} A.J. Keane and R.K. Barrett, \textit{The conformal group SO(4,2) and Robertson-Walker spacetimes}, Class. Quantum Grav. \textbf{17}, 201 (2000).
\bibitem{IKM} M. Iihoshi, S.V. Ketov and A. Morishita, \textit{Conformally flat FRW metrics}, Prog. Theor. Phys. \textbf{118}, 475 (2007).
\bibitem{Co} F.I. Cooperstock, \textit{The intteraction between electromagnetic and gravitational waves}, Ann. Phys. \textbf{47}, 173 (1968).
\bibitem{Cr} A.M. Cruise, \textit{An intteraction between electromagnetic and gravitational waves}, Mon. Not. R. Astron. Soc. \textbf{204}, 485 (1983).
\bibitem{AG} G.A. Alekseev and J.B. Griffiths, \textit{Collision of plane gravitational and electromagnetic waves in a Minkowski background: solution of the characteristic initial value problem}, Class. Quantum Grav. \textbf{21}, 5623 (2004)
\bibitem{F} V. Faraoni, \textit{The rotation of polarisation by gravitational waves}, New Astronomy \textbf{13}, 178 (2008).
\bibitem{BH} C. Barrabes and P.A. Hogan, \textit{Interaction of gravitational waves with magnetic and ellectric fields}, \textbf{81}, 064024 (2010).
\bibitem{LL} L.D. Landau and E.M. Lifshitz, \textit{A course in theoretical physics: Vol.~I} (Butterworth-Heinemann, Amsterdam, 1976).
\end{thebibliography}
\end{document}